\documentclass[10pt, conference, letterpaper]{IEEEtran}
\IEEEoverridecommandlockouts
\usepackage[style=ieee, doi=false, url=false, isbn=false, giveninits=true, backend=biber, sorting=none, sortcites=true, date=year]{biblatex}
\usepackage{amsmath,amssymb,amsfonts}
\usepackage{graphicx}
\usepackage{textcomp}
\usepackage{xcolor}
\definecolor{Honeydew}{RGB}{240, 255, 240}
\usepackage{url}

\usepackage{booktabs}
\usepackage{multirow}
\usepackage{makecell}
\usepackage{algorithmicx}%
\usepackage{algorithm}
\usepackage{subfigure}
\usepackage{colortbl}
\usepackage{algpseudocode}%
\usepackage{hyperref}
\usepackage{bm}
\usepackage[capitalize]{cleveref}
\Crefname{section}{Section}{Sections}
\Crefname{table}{Table}{Tables}
\usepackage[acronym]{glossaries}
\newacronym{e2e}{E2E}{End-to-End}
\newacronym{jscc}{JSCC}{Joint Source-Channel Coding}
\newacronym{tscc}{TSCC}{Task-oriented Source-Channel Coding}
\newacronym{snr}{SNR}{Signal-to-Noise Ratio}
\newacronym{rf}{RF}{Radio Frequency}
\newacronym{dnn}{DNN}{Deep Neural Network}
\newacronym{ber}{BER}{Bit-Error Rate}
\newacronym{kl}{KL}{Kullback-Leibler}
\newacronym{vae}{VAE}{Variational Autoencoder}
\newacronym{cvae}{CVAE}{Conditional Variational Autoencoder}
\newacronym{awgn}{AWGN}{Additive White Gaussian Noise}
\newacronym{tcp}{TGCP}{Trajectory-guided Control Prediction}
\newacronym{bpp}{BPP}{bits per pixel}
\newacronym{psnr}{PSNR}{Peak Signal-to-Noise Ratio}
\newacronym{msssim}{MS-SSIM}{Multi-Scale Structural Similarity}
\newacronym{ssim}{SSIM}{Structural Similarity}
\newacronym{v2x}{V2X}{Vehicle-to-Everything}
\newacronym{fid}{FID}{Fréchet Inception Distance}
\newacronym{drl}{DRL}{Deep Reinforcement Learning}
\newacronym{ppo}{PPO}{Proximal Policy Optimization}
\newacronym{mse}{MSE}{Mean Squared Error}
\newacronym{ugv}{UGV}{Unmanned Ground Vehicle}
\newacronym{svd}{SVD}{Singular Value Decomposition}
\newacronym{lsa}{LSA}{Latent Semantic Analysis}
\newacronym{ai}{AI}{Artificial Intelligence}
\newacronym{embb}{eMBB}{enhanced Mobile Broadband}
\newacronym{mmtc}{mMTC}{massive Machine-Type Communications}
\newacronym{urllc}{URLLC}{Ultra-Reliable Low-Latency Communications}
\newacronym{cc}{C\&C}{control and command}
\newacronym{tosa}{TOSA}{task-oriented semantics-aware}
\newacronym{aoi}{AoI}{Age of Information}
\newacronym{qos}{QoS}{Quality of Service}
\newacronym{vqvae}{VQ-VAE}{Quantized Variational Autoencoders}
\newacronym{ib}{IB}{information bottleneck}
\newacronym{uav}{UAV}{Unmanned Aerial Vehicle}
\newacronym{mec}{MEC}{Mobile Edge Computing}
\newacronym{qoe}{QoE}{Quality of Experience}
\newacronym{dvr}{DVR}{Deadline Violation Ratio}
\newacronym{ddpg}{DDPG}{Deep Deterministic Policy Gradient}
\newacronym{kpi}{KPI}{Key Performance Indicators}

\DeclareSourcemap{
  \maps[datatype=bibtex]{
    \map{
      \step[fieldset=primaryClass, null]
      \step[fieldset=publisher, null]
      \step[fieldset=editor, null]
    }
  }
}
\def\BibTeX{{\rm B\kern-.05em{\sc i\kern-.025em b}\kern-.08em
T\kern-.1667em\lower.7ex\hbox{E}\kern-.125emX}}

\begin{document}

\title{Task-Oriented Edge-Assisted Cooperative Data Compression, Communications and Computing for UGV-Enhanced Warehouse Logistics\\}
%{\footnotesize \textsuperscript{*}Note: Sub-titles are not captured in Xplore and should not be used}
\author
{
Jiaming Yang\textsuperscript{1}, 
Zhen Meng\textsuperscript{1}, 
Xiangmin Xu\textsuperscript{1}, 
Kan Chen\textsuperscript{1}, 
Emma Liying Li\textsuperscript{1}, 
Philip G. Zhao\textsuperscript{2} \\

\textit{\textsuperscript{1}School of Computer Science, University of Glasgow, UK} \\

\textit{\textsuperscript{2}Department of Computer Science, University of Manchester, UK} \\

\textit{\{j.yang.9, x.xu.1, k.chen.1\}@research.gla.ac.uk},\\ 

\textit{\{zhen.meng, liying.li\}@glasgow.ac.uk, philip.zhao@manchester.ac.uk}
}

\maketitle
\begin{abstract}

This paper explores the growing need for task-oriented communications in warehouse logistics, where traditional communication Key Performance Indicators (KPIs)—such as latency, reliability, and throughput—often do not fully meet task requirements. As the complexity of data flow management in large-scale device networks increases, there is also a pressing need for innovative cross-system designs that balance data compression, communication, and computation. To address these challenges, we propose a task-oriented, edge-assisted framework for cooperative data compression, communication, and computing in Unmanned Ground Vehicle (UGV)-enhanced warehouse logistics. In this framework, two UGVs collaborate to transport cargo, with control functions—navigation for the front UGV and following/conveyance for the rear UGV—offloaded to the edge server to accommodate their limited on-board computing resources. We develop a Deep Reinforcement Learning (DRL)-based two-stage point cloud data compression algorithm that dynamically and collaboratively adjusts compression ratios according to task requirements, significantly reducing communication overhead. System-level simulations of our UGV logistics prototype demonstrate the framework's effectiveness and its potential for swift real-world implementation.

% The wave of shifting from traditional to task-oriented communications is radiating vertical industries, aiming at more finely segmented mission-critical tasks such as autonomous driving, warehousing and logistics. However, complex information flow pipelines in these applications often require cross-system designs that balance sampling, communication, and computation, a problem that is even more challenging when there are large-scale device deployments. In this paper, we present a design framework for task-oriented edge computing-based cooperative data compression, communication, and control, using a collaborative handling through automated guided vehicle scenario in a warehouse as an example. We propose a ... to dynamically compress the point cloud data for different task tendencies of the front and rear vehicles. We built a prototype and tested it in a high simulation warehouse environment with different baselines, and the experimental results validate that our proposal can significantly reduce the communication overhead and can be quickly deployed to the real world.

\end{abstract}
\glsresetall

% Should be deleted before submission
%\pagecolor{Honeydew}

\section{Introduction}

Warehousing plays an essential role within the supply chain by serving as a crucial hub for systematic storage, meticulous management, and streamlined distribution of cargos. It serves as a strategic node where inventory is carefully organized, monitored, and prepared for timely delivery to meet consumer demands and operational requirements within various industries~\cite{richards2017warehouse}. Modern warehouses are progressively incorporating advanced technologies such as \gls{ai} and robotics to facilitate intelligent automation of tasks~\cite{nvidia_warehouse}. However, the deployment of these technologies necessitates extensive data transmission (e.g. radar, camera, and GPS), real-time responsiveness (e.g. decision making for autonomous navigation), and substantial machine-type communications (e.g. multi-\gls{ugv} coordination). Despite these requirements aligning with the three typical 5G services defined by 3GPP—\gls{embb}, \gls{mmtc}, and \gls{urllc}-they still do not fully meet the stringent demands of warehouse applications~\cite{3GPP, 9509294}. This is particularly true as more complex tasks emerge in the foreseeable future~\cite{nvidia_warehouse}. Due to the limited communication capabilities and computational load of \gls{ugv} in warehouse logistics scenarios, edge computing has emerged as a potential solution~\cite{8016573, dinh2013survey}. By offloading computational tasks to edge servers, \glspl{ugv} can achieve more powerful intelligence, reduced communication, and computational resource overhead, and significantly faster response and decision-making capabilities~\cite{8493155}.

\begin{figure}[t]
    \begin{center}
    \includegraphics[width=\linewidth]{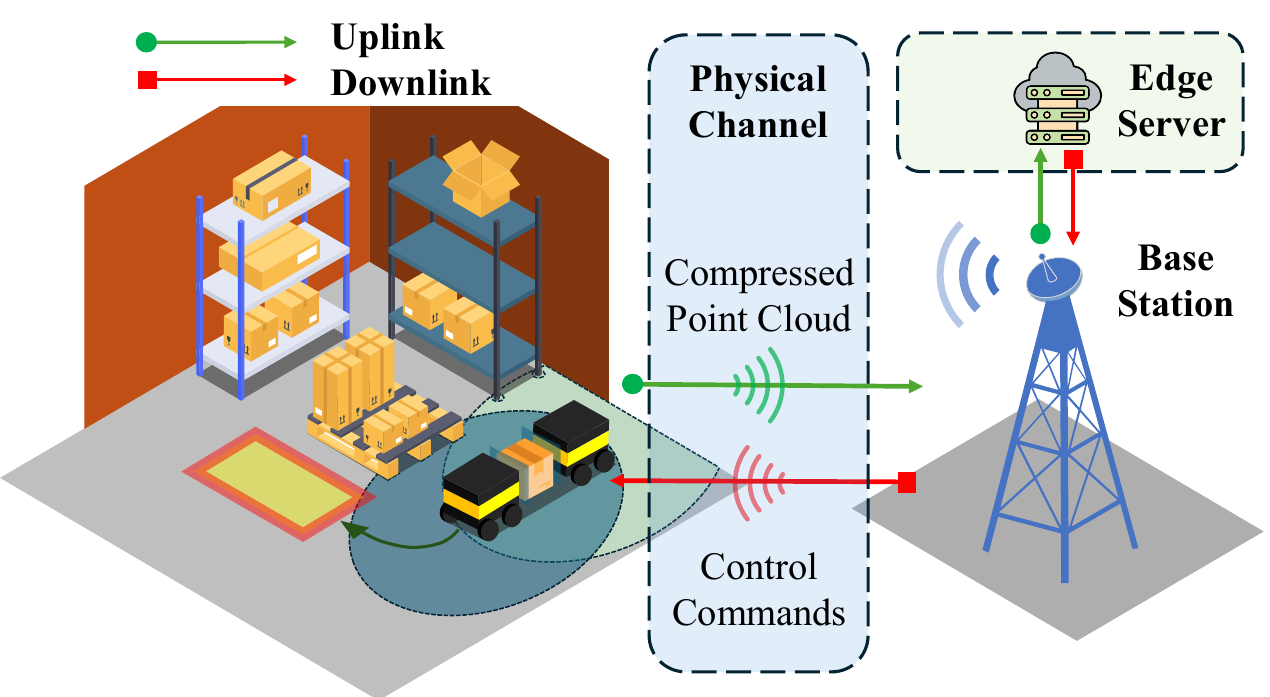}
    \end{center}
       \caption{Edge computing enabled autonomous driving.}
    \label{fig_general_idea}
\end{figure}

From a machine's perspective, achieving a perfectly accurate reconstruction is deemed sufficient but unnecessary for task completion, as it often contains redundant information. The ultimate objective of transmission is to facilitate the completion of subsequent tasks. Based on that, task-oriented communication is gradually gaining attention~\cite{10554663}. It follows Shannon's second and third ideas, moving away from the bit-centric communication principles towards a communication design approach that focuses on meeting semantic and task requirements~\cite{9955525}. This communication design approach is also gradually being applied in wide-ranging applications, including robotics~\cite{10370739}, autonomous driving~\cite{yufeng}, and metaverse~\cite{10422886} scenarios.

% However, the industry faces several challenges: 1) labor shortages impacting efficiency \cite{nvidia_warehouse}, 2) rising operational costs due to automation investments \cite{nvidia_deep}, 3) space utilization issues driven by e-commerce growth, 4) the complexity of integrating new technologies \cite{nvidia_isaac}, and 5) ensuring the quality and relevance of synthetic data used for training AI models \cite{nvidia_synthetic}.

%\Zhen{Talk a little bit about warehousing, you can get some inspiration and keywords from NVIDIA advertisement of isaac sim for warehousing. Second, talk about the challenges of warehousing currently facing.}
%AI,warehouse:  https://www.nvidia.com/en-gb/industries/retail/warehouse-logistics/
%DEEP,WAREHOUSE:   https://developer.nvidia.com/blog/deep-learning-in-robotic-automation-and-warehouse-logistics/
%AI,ISAACSIM:   https://developer.nvidia.com/blog/expedite-the-development-testing-and-training-of-ai-robots-with-isaac-sim/
%data:  https://www.nvidia.com/en-gb/use-cases/synthetic-data/

However, task-oriented communications still face bottlenecks in \gls{ugv}-enhanced warehouse logistic applications, including: 1) Gap between traditional Communication \glspl{kpi} and Task-oriented KPIs: The more detailed and intricate scenarios within warehouse applications involve complex interactions with their surroundings. As a result, connecting traditional communication \glspl{kpi} (e.g. latency, throughput, and reliability) to the success of tasks becomes difficult. This difficulty arises because these complex interactions cannot be easily represented or measured using straightforward loss functions or performance metrics. 2) Seamless Data Sharing and Interpretation: Different submodules in warehouse logistics systems must seamlessly share and interpret data to perform cooperative tasks effectively. Therefore, managing these information pipelines requires cross-system designs that balance sampling, communication, and computation demands. This balance is crucial given the vast amounts of data generated by numerous devices and the need for real-time processing. 3) Scalability: As the number of \glspl{ugv} increases, the efficient deployment of the \gls{ai} algorithm becomes challenging. Therefore, how to handle complex information flows in large-scale \gls{ugv} deployments in warehouses presents significant challenges.
% However, task-oriented communications still face bottlenecks in \gls{ugv}-assisted warehouse applications, which includes: 1) The gap between the traditional communication KPIs and tasks-oriented KPIs in terms of warehouse applications -- More granular and complex application scenarios defined in the warehouse scenarios: Some Complex interactions in the external environment that cannot be modeled with explicit loss function of performance evaluated metrics.  2) Different systems must share and interpret data seamlessly to perform cooperative tasks effectively. The complexity of managing these information pipelines necessitates cross-system designs that balance sampling, communication, and computation demands. This balance becomes even more critical with the vast amounts of data generated by numerous devices and the requirement for real-time processing. 3) Scalability: how to handle complex information flows in large-scale deployments for example like \gls{ugv} in warehouse presenting significant challenges. 

To address these challenges, we propose a task-oriented edge-assisted cooperative data compression, communications, and computing framework for \gls{ugv}-enhanced warehouse logistics. Specifically, we take typical warehouse logistics collaboration tasks, where two \glspl{ugv} convey cargo boxes cooperatively to target positions, as example applications as the showcase~\cite{8927467}. Constrained by the on-board computing capacities of \glspl{ugv}, we offload the two controllers of the \glspl{ugv}; the front \gls{ugv} for navigation and rear \gls{ugv} for collaborated conveyance. To save communication overhead, we design a \gls{drl}-based two-stage point cloud data compression algorithm deployed on the edge server, where the compression ratios of the two \glspl{ugv} are dynamically and collaboratively determined based on the task requirements. 
% In particular, we develop a prototype and simulate its algorithm in a warehouse environment alongside various baselines. The experimental results demonstrate that our proposal significantly reduces communication overhead under different channel conditions and can be quickly implemented in real-world settings.
% my research employs Latent Semantic Analysis for adaptive compression, simplifying the process while maintaining efficient data transmission. Unlike the approach of pre-training multiple models at different compression rates, LSA enables adaptive compression that dynamically adjusts based on the data, reducing complexity and ensuring that essential information is retained. This method improves the overall efficiency and robustness of data transmission, aligning with the broader goal of optimizing communication strategies under varying conditions and supporting advancements in semantic and effective communication for complex robotic operations. In addition, this framework is demonstrated through a scenario involving \glspl{ugv} in a warehouse environment. Our approach leverages a reinforcement learning (RL) network to dynamically compress point cloud data based on the specific task requirements of the front and rear ugvs, thereby optimizing communication efficiency and data handling.
The major contributions of this work are summarized as follows:
\begin{itemize}
\item We propose a task-oriented edge-assisted framework for \gls{ugv}-enhanced warehouse logistics, where the data compression, communications, and computing of front and rear \glspl{ugv} are jointly considered for accomplishing collaborative cargo box conveyance and parking tasks. The functions of two controllers of two \glspl{ugv} are offloaded to the edge server, and the compression rate of two \glspl{ugv} are jointly and dynamically determined based on the channel conditions and task requirements. 
\item We proposed a two-stage \gls{drl}-based point cloud data compression algorithm to save the communication overhead, simultaneously exploiting the effectiveness of neural networks and the stability of model-based. Specifically, the data is first initially compressed by a data-based neural network compressor, then undergoes model-based compression and the compression length is dynamically decided by a \gls{drl}-based agent. 
\item We developed a prototype and verified the robustness of the proposed algorithm in a warehouse environment alongside various baselines, where the Nvidia Issac Sim platform is used. The experimental results demonstrate that our proposal significantly reduces communication overhead under different channel conditions which can be quickly deployed in the real world.

\end{itemize}

% \begin{itemize}
%     \item Develop a scalable and efficient control framework that can be applied to autonomous warehouse operations. This framework should enhance the overall capabilities of robotic systems in logistics and delivery tasks, contributing to improved efficiency and productivity in warehouse operations. Address the limitations posed by limited communication resources within the warehouse. Ensure reliable and continuous sharing of sensor data and coordination signals between the two UGVs, even in the presence of wireless interference and bandwidth constraints.   
%     % framework jscc, imitation learning? 
%     \item Develop advanced navigational algorithms to ensure the UGVs can safely and efficiently navigate the complex warehouse environment. These algorithms must dynamically adjust to new information and environmental changes, avoiding collisions with stationary and dynamic obstacles.
%     \item  Design and implement a robust control system that enables two Jackal UGVs to work together effectively in transporting parcels within a warehouse. One ugv will lead while the other follows, securely clamping and carrying the cargo between them.

% \end{itemize}
\section{Related work}
Significant contributions have been made in the existing literature to improve the intelligence of different \glspl{ugv} and reduce communication overhead. By processing data closer and offloading tasks to where it is generated, edge computing reduces latency and the amount of data transmitted to central servers, conserving bandwidth and improving response times~\cite{he2024qoemaximizationmultipleuavassistedmultiaccess, 9910575, 10007839, 9406800}. 
In~\cite{9910575}, the authors present cloud control of \glspl{ugv} in future factories. By optimizing the coding rate threshold through control parameter adjustments, the study shows that communication-control co-design reduces coding rate requirements and wireless resource consumption, improving system stability and increasing the number of admissible \glspl{ugv}. 
In~\cite{he2024qoemaximizationmultipleuavassistedmultiaccess}, the authors address disaster scenarios by proposing \gls{uav}-assisted \gls{mec} as an alternative to damaged terrestrial infrastructure. They introduce a hierarchical architecture and an online optimization approach (OJTRTA) using game theory and convex optimization to improve \gls{qoe} and manage \gls{uav} resources. In~\cite{10007839}, the authors address computation offloading in \gls{mec} systems with task dependencies by proposing a scheme that uses task migration and merging to minimize \gls{dvr}. They introduce a multi-priority task sequencing algorithm and a \gls{ddpg}-based learning approach, achieving a $60.34\%$ to $70.3\%$ reduction in \gls{dvr} compared to existing methods.

On the other hand, task-oriented communications have been applied in widespread applications scenarios including text~\cite{Xie2021DeepSC}, image~\cite{Zhang2024Optimization}, point cloud data transmission~\cite{9414831}, and control~\cite{10164147} tasks. In~\cite{10164147}, the authors propose a \gls{tosa} communication framework for \gls{uav} \gls{cc} transmissions to meet stringent \gls{qos} requirements. They define information value based on similarity and \gls{aoi} and use a \gls{drl} algorithm to maximize \gls{tosa} information 
The work in~\cite{Talli2023Dynamic} addresses optimizing communication for 5G and beyond in robotic swarms and industrial remote control systems. It proposes a dual approach using \gls{drl} and Vector \gls{vqvae} to dynamically adjust data transmission for better accuracy and compression on the CartPole problem~\cite{yufeng}. The authors in~\cite{9606667} further extend the task-oriented communications framework to the edge server, where the feature extraction, source coding, and channel coding are jointly designed and guided by \gls{ib} theory. The results show that it can reduce communication overhead through a sparsity-inducing distribution and adapt to dynamic channel conditions with variable-length encoding. In addition, there has been a lot of existing work focusing on task-oriented communication in multi-user situations and semantic communication in applications~\cite{10570351, 10183796, 10520522, 9837474, 9830752}.  In~\cite{10183796}, the author proposes a task-oriented communication-based framework for Multi-Agent systems, aiming to support efficient cooperation among agents. However, intrinsic connections between \glspl{ugv} in warehouse logistics, especially in the context of relationships with collaborative tasks and conditions of resource competition, remain an open issue.

\begin{figure*}[t]
    \begin{center}
    \includegraphics[width=\linewidth]{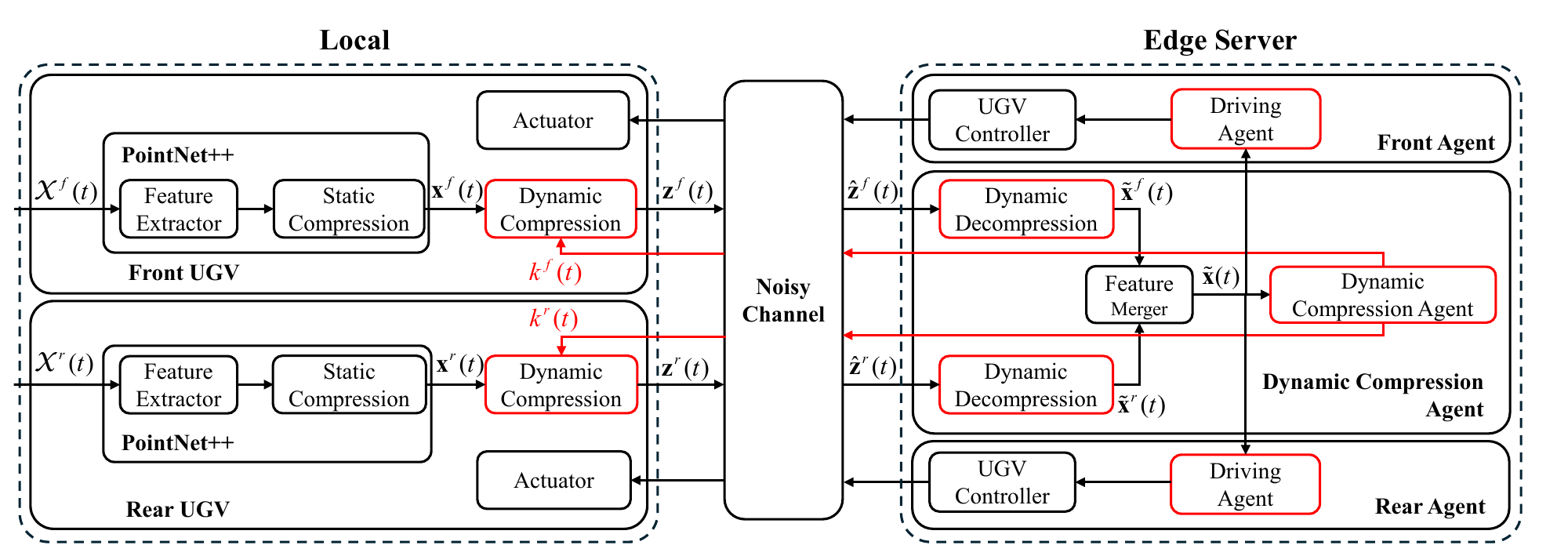}
    \end{center}
       \caption{Integrated Feature Extraction, Reinforcement Learning, and Data Compression in a Cooperative Semantic Communication System.}
    \label{fig:framework}
\end{figure*}

\section{System Model}
% \Zhen{Clearly defined the overall framework and A, B, C each sub-module}

\subsection{Overview}
As shown in Fig.~\ref{fig_general_idea}, two \glspl{ugv} convey a cargo box collaboratively to park in a certain place. The function of the front \gls{ugv} is to navigate to the desired parking place, and the role of the rear \gls{ugv} is to follow and cooperate with the front \gls{ugv} in conveying the cargo box. They both fulfill their tasks based on the captured 3-D point-cloud data from the onboard LiDAR. Constrained by their onboard computing capacities, we offload the two controllers of two \glspl{ugv} to the edge server. That is, the captured 3-D point cloud data is first transmitted by the two \glspl{ugv} via the uplink wireless channel. After receiving and processing by the edge server, the control commands of two \glspl{ugv} are generated and transmitted back via the downlink channel. In addition, to save communication overhead, we deploy two a \gls{drl}-based two-stage point cloud data compressor on two \glspl{ugv}, where the compression rate of two \glspl{ugv} is dynamically and collaboratively determined based on the task requirements by another \gls{drl} agent deployed on the edge server.

% As shown in Fig.~\ref{fig:framework}, the framework consists of a collaborative pair of \glspl{ugv} and an edge server. The two \glspl{ugv} align the boards mounted on their front and back, clamp the parcels and transport them to designated locations. The front \gls{ugv} handles parking tasks and point cloud data compression, while the rear \gls{ugv} is responsible for tracking and following. The edge server processes and transmits commands to both \glspl{ugv}. Both \glspl{ugv} and the edge server communicate with each other via wireless channels. The communication between the \glspl{ugv} is lossless, ensuring reliable point cloud data transmission. However, the communication between the \glspl{ugv} and the edge server is lossy, which may result in occasional data loss. Due to communication latency and packet loss when transmitting, there are discrepancies in the collaboration between the \glspl{ugv}. When the \glspl{ugv} work together to complete tasks, the system's edge server utilizes three \gls{drl} agents to achieve 1) Automate the parking of the front \gls{ugv}. 2) Ensure the rear \gls{ugv} follows the front \gls{ugv}, collaborating effectively to complete tasks. 3) Optimize the performance of the collaborative work, reducing latency between the \glspl{ugv} and the edge server.

% The two \glspl{ugv} perform their work together, where surrounding environment is scanned by Lidar and converted into point cloud data. 

\subsection{Two-Stages Compression}
As shown in Fig.~\ref{fig:framework}, two \glspl{ugv} capture the 3-D point cloud data of the front, ${\mathcal{X}}^f(t) = \{x_1, x_2, ..., x_m\}, x_m \in \mathbb{R}^d$ and the rear, ${\mathcal{X}}^r(t) = \{x_1, x_2, ..., x_n\}, x_n \in \mathbb{R}^d$ by the on-board lidar at a certain sensing rate in the $t$-th time slot, where $d$ is the dimension.
% which are denoted by 
% \begin{align}\label{eq: point_cloud}
% {\mathcal{X}_t}^f = \{x_1, x_2, ..., x_m\}, \ \ x_t \in \mathbb{R}^d, \\
% {\mathcal{X}_t}^r = \{x_1, x_2, ..., x_n\}, \ \ x_i \in \mathbb{R}^d
% \end{align}
% \gls{ugv} and the rear \gls{ugv} are unordered point sets, $\mathcal{X}^f, \mathcal{X}^r \rightarrow \mathbb{R}$, which are denoted by 
% two \glspl{ugv} scanned the environment separately, the sampled point cloud data for the front \gls{ugv} and the rear \gls{ugv} are unordered point sets, $\mathcal{X}^f, \mathcal{X}^r \rightarrow \mathbb{R}$, which are denoted by 
% \begin{align}\label{eq: point_cloud}
% {\mathcal{X}_t}^f = \{x_1, x_2, ..., x_m\}, \ \ x_t \in \mathbb{R}^d, \\
% {\mathcal{X}_t}^r = \{x_1, x_2, ..., x_n\}, \ \ x_i \in \mathbb{R}^d
% \end{align}
To compress the captured 3-D point cloud data to reduce the communication overhead, we introduce a two-stage \gls{drl}-based point cloud data compression module. Here we propose to use \textit{PointNet++} as the static compressor for its simple structure, widespread application, and suitability for complex scene compression~\cite{NIPS2017_d8bf84be}. Specifically, firstly, in the $t$-th time slot, ${\mathcal{X}}^f(t)$ and ${\mathcal{X}}^r(t)$ passed through the feature extractor part of \textit{PointNet++}, which is a hierarchical neural network expressed by
% PointNet++ for the preliminary compression of the scanned point cloud data due to its simple structure, widespread application, and suitability for complex scenes~\cite{}. Before compression, the raw data of point cloud requires a complex neural network for feature extraction. 
% The point cloud data will be transmitted to the edge server for processing, but it needs to be compressed before transmission to conserve communication resources. Since directly compressing the data to an optimal size is challenging, the point cloud data will first processed by preliminary compression, and followed by \gls{drl} based dynamic compression. 
% We introduced PointNet++ for the preliminary compression of the scanned point cloud data due to its simple structure, widespread application, and suitability for complex scenes~\cite{}. Before compression, the raw data of point cloud requires a complex neural network for feature extraction. In the $t$-th time slot, the process of feature extraction is expressed by
\begin{align}\label{eq: feature extractor}
\bar{\mathcal{X}}^f(t) = f_{e^1}(\mathcal{X}^f(t),\theta_{e^1}), \\
\bar{\mathcal{X}}^r(t) = f_{e^2}(\mathcal{X}^r(t),\theta_{e^2}),
\end{align}
where $\theta_{e^1}, \theta_{e^2}$ denote the parameters of the two feature extractors. Secondly, $\bar{\mathcal{X}}^f(t)$ and $\bar{\mathcal{X}}^r(t)$ are compressed to the vector ${\bf{x}}^f(t)$ and ${\bf{x}}^r(t)$ with a fixed latent space by passing trough two fully connected neural networks, separately, which is expressed by,
% Then $\bar{\mathcal{X}}_t^f$ and $\bar{\mathcal{X}}_t^r$ are further 
% Then, followed by a hierarchical grouping of points and progressively maps the compressed vector, ${\bf{x}}_t^f$ and ${\bf{x}}_t^r$ with fixed dimensions, which is denoted by 
\begin{align}\label{eq: static compressor}
{\bf{x}}^f(t) = f_{s^1}(\bar{\mathcal{X}}^f(t),\theta_{s^1}),\\
{\bf{x}}^r(t) = f_{s^2}(\bar{\mathcal{X}}^r(t),\theta_{s^2}),
\end{align}
where $f_{s^1}(\cdot,\theta_{s^1}), f_{s^2}(\cdot,\theta_{s^2})$ represent the dimensional reduction neural networks and $\theta_{s^1}, \theta_{s^2}$ are the corresponding parameters. 

For the second-stage model-based data compressor, we propose to use \gls{lsa} for its simplicity, robustness, and no need for extensive hyper-parameter tuning~\cite{lsathomas}. It works by decomposing a term-document matrix using \gls{svd} to reveal and map the underlying semantic structure into a lower-dimensional space. Specifically, in the $t$-th time slot, ${\bf{x}}^f(t)$, and ${\bf{x}}^r(t)$ are firstly decomposed by \gls{svd} separately, which is denoted by
\begin{align}
    U_f(t) \Sigma_f(t) V_{f}^T(t) = {\bf{x}}^f(t), \label{eq: compress_1}\\
    U_r(t) \Sigma_r(t) V_{r}^T(t) = {\bf{x}}^r(t), \label{eq: compress_2}
\end{align}
where $U_f(t) \in \mathbb{R}^{1 \times 1}$ and $U_r(t) \in \mathbb{R}^{1 \times 1}$ are the left singular vectors for the front and rear \glspl{ugv} respectively, $\Sigma_f(t) \in \mathbb{R}^{1 \times n}$ and $\Sigma_r(t) \in \mathbb{R}^{1 \times n}$ are the diagonal matrices for the front and rear \glspl{ugv} with non-negative singular values in descending order, and $V_f^T(t) \in \mathbb{R}^{n \times n}$ and $V_r^T(t) \in \mathbb{R}^{n \times n}$ are the right singular vectors for the front and rear \glspl{ugv}, respectively. Secondly, by selecting the top $k^f$, $k^r$ singular values from $\Sigma_f(t)$, $\Sigma_r(t)$, and their corresponding vectors from $U_f(t)$, $U_r(t)$, $V_{f}^T(t)$ and $V_{r}^T(t)$, we achieve the dimensionality reduction by operating the inverse of SVD,
\begin{align}
    {\bf{z}}_f(t) = U_f'(t) \Sigma_f'(t) V_f'^T(t), \ \  \label{eq: compress_3}\\ 
    {\bf{z}}_r(t) = U_r'(t) \Sigma_r'(t) V_r'^T(t), \ \ \label{eq: compress_4}
\end{align}
where $U_f'(t), U_r'(t) \in \mathbb{R}^{1\times 1}$,  $\Sigma'_f(t) \in \mathbb{R}^{1\times k^f(t)}$ and $\Sigma'_r(t) \in \mathbb{R}^{1\times k^r(t)}$ are the compressed diagonal matrices, $V_f'^T(t) \in \mathbb{R}^{k^f(t)\times k^f(t)}$ and $V_r'^T(t) \in \mathbb{R}^{k^r(t)\times k^r(t)}$ are the compressed right singular vectors, and $k^f(t), k^r(t) \in \mathcal{Z}$ are the dynamic compression parameters. Thus, by synthesizing the information from results of (\ref{eq: compress_1}), (\ref{eq: compress_2}), (\ref{eq: compress_3}), (\ref{eq: compress_4}), the dynamic compression in the $t$-th time slot is expressed by
\begin{align}
\label{dynamiccompression}
\{{\bf{z}}^f(t),{\bf{z}}^r(t)\} = f_d(\{{\bf{x}}^f(t), {\bf{x}}^r(t)\},\theta_d,\{k^f(t),k^r(t)\}),
\end{align}
where $f_d(\cdot,\theta_d)$ is the process of dynamic compressor, $\theta_d$ represents the parameters of the compressor. To dynamically change the compression ratio, we deploy a \gls{drl} agent at the edge server to dynamically decide the value of $k^f(t)$ and $k^r(t)$ and send them back to each \glspl{ugv} via wireless channel. The details will be provided in the following Section. After the two-stage compression, the compression ratio of ${\bf{z}}^f(t)$ and ${\bf{z}}^r(t)$ compared to original cloud point data $\mathcal{X}^f(t)$ and $\mathcal{X}^r(t)$  is given as
\begin{align}
    \rho^f(t) = \frac{k^f(t)}{m \cdot d},
\end{align}
\begin{align}
    \rho^r(t) = \frac{k^r(t)}{n \cdot d}.
\end{align}
% This process retains the most significant semantic structures while reducing noise and dimensionality.    
% An additional \gls{drl} setup is utilized to dynamically adjust the compression ratio of the secondary compression by adjusting the value of $k$. This adaptive mechanism aims to achieve higher compression rates while ensuring effective control, optimizing data transmission efficiency without compromising operational effectiveness.
% To reduce dimensionality, \gls{lsa} selects the top $k$ singular values and their corresponding vectors, resulting in the compressed matrix, which is denoted by
\begin{algorithm}[t]
\caption{Two-Phase \gls{drl} Training}
\label{algorithm1}
\begin{algorithmic}[1]
\State \textbf{Input}: Initialize the navigation policy $ \pi_{\theta_1}$, cooperative conveyance policy $\pi_{\theta_2}$, compression policy $\pi_{\theta_3}$, the parameters of neural network $\theta_1, \theta_2$ and $\theta_3$, the positions of the \glspl{ugv} and the cargo box ${\bf{p}}^f_0, {\bf{p}}^r_0$ and ${\bf{p}}^c_0$, and the total training steps $T_t$.

\State \textbf{Phase 1}: Training policies $\pi_{\theta_1}$ and $\pi_{\theta_2}$.
\For{$t = 1, 2, ... T_t$}
    \State Sample $\mathcal{X}_f(t)$ and $\mathcal{X}_r(t)$ from the front and rear
    \Statex \quad \; \glspl{ugv}.
    \State ${\bf{x}}_f(t) \gets f_{s^1} \left( f_{e^1}(\mathcal{X}^f(t),\theta_{e^1}), \theta_{s^1})\right.$, 
    \Statex \quad \; ${\bf{x}}_r(t) \gets f_{s^2} \left( f_{e^2}(\mathcal{X}^r(t),\theta_{e^2}), \theta_{s^2})\right.$.
    % \Statex \quad \; Compress $\mathcal{X}_f(t)$ and $\mathcal{X}_r(t)$ into feature vectors ${\bf{x}}_f(t)$ 
    % \Statex \quad \; and ${\bf{x}}_r(t)$ with the proposed first-stage compression 
    % \Statex \quad \; algorithm.
    \State ${\bf{x}}(t) \gets [{\bf{x}}_f(t), {\bf{x}}_r(t)]$.
     \State ${\bf{s}}^{[1]}_t \gets {\bf{x}}(t)$, ${\bf{s}}^{[2]}_t \gets {\bf{x}}(t)$.  
    \State ${\bf{a}}^{[1]}_t \gets \pi_{\theta_1}({\bf{s}}^{[1]}_t)$, ${\bf{a}}^{[2]}_t \gets \pi_{\theta_2}({\bf{s}}^{[2]}_t)$. 
    \State Calculate $r^{[1]}({\bf{s}}^{[1]}_t,{\bf{a}}^{[1]}_t)$ with (\ref{r1}), 
    \Statex \quad \; and $r^{[2]}({\bf{s}}^{[2]}_t,{\bf{a}}^{[2]}_t)$ with (\ref{r2}).
    \State Collect tuple $\{\tilde{\bf{x}}_t, {\bf{a}}^{[1]}_t, {\bf{a}}^{[2]}_t, r^{[1]}({\bf{s}}_t^{[1]}, {\bf{a}}_t^{[1]}), r^{[2]}({\bf{s}}_t^{[2]}, {\bf{a}}_t^{[2]})\}$
    \If {Episode ends}:
        \State Calculate the Q-Values $Q^{\pi_{\theta_1}}$ and $Q^{\pi_{\theta_2}}$ with
        \Statex \quad \; \quad \;  $Q^{\pi_{\theta}}({{\bf{s}}_t}, {{\bf{a}}_t}) \gets {\mathop{\mathbb{E}}}[\sum_{t = 0}^\infty  {{ \gamma ^t}}r({\bf{s}}_t, {\bf{a}}_t) \mid \pi_{\theta}]$.
        \State Train $\theta_1$ and $\theta_2$ to maximize $Q^{\pi_{\theta_1}}$ and $Q^{\pi_{\theta_2}}$.
        % \State Calculate $Q^{\pi_{\theta_1}}$ and $Q^{\pi_{\theta_2}}$ with (\ref{qvalue}).
        % \State Update $\theta_1$ and $\theta_2$ with (\ref{pi1}) and (\ref{pi2}).
    \EndIf
    % \State Maximize $Q^{\pi_{\theta_1}}$ and $Q^{\pi_{\theta_2}}$, respectively.
\EndFor
\State \textbf{Output}: Optimal policies $\pi^*_{\theta_1}$ and $\pi^*_{\theta_2}$.

\State \textbf{Phase 2}: Training policy $\pi_{\theta_3}$.
\For{$t = 1, 2, ... T_t$}
    \State Sample and compress to get ${\bf{x}}^f(t)$ and ${\bf{x}}^r(t)$ 
    \Statex \quad \; with Step 4 to Step 5.
    \State $\{{\bf{z}}^f(t),{\bf{z}}^r(t)\} \gets f_d(\{{\bf{x}}^f(t), {\bf{x}}^r(t)\},\theta_d,\{k^f(t),k^r(t)\}).$ 
    \Statex \quad \; Dynamic compression with (\ref{dynamiccompression}).
    \State Calculate $T_{up}$ with (\ref{tup}).
    \State Decompress $\tilde{\bf{x}}(t)$ with (\ref{decompress}).
    \State ${\bf{s}}^{[1]}_t \gets \tilde{\bf{x}}(t)$, ${\bf{s}}^{[2]}_t \gets \tilde{\bf{x}}(t)$,
    ${\bf{s}}^{[3]}_t \gets [\tilde{\bf{x}}(t), g^f(t), g^r(t)]$
    
    \State ${\bf{a}}^{[1]}_t \gets \pi^*_{\theta_1}({\bf{s}}^{[1]}_t)$, ${\bf{a}}^{[2]}_t \gets \pi^*_{\theta_2}({\bf{s}}^{[2]}_t)$. 
    \State ${\bf{a}}^{[3]}_t \gets \pi_{\theta_3}({\bf{s}}^{[3]}_t)$.
    \State Calculate $r^{[3]}({\bf{s}}^{[3]}_t,{\bf{a}}^{[3]}_t)$ with (\ref{r3}).
    \State Collect tuple $\{\tilde{\bf{x}}_t, g^f(t), g^r(t), {\bf{a}}^{[3]}_t, r^{[3]}({\bf{s}}_t^{[3]}, {\bf{a}}_t^{[3]})\}$.
    \If {Episode ends}:
        \State Calculate $Q^{\pi_{\theta_3}}$ with \Statex \quad \; \quad \;  $Q^{\pi_{\theta}}({{\bf{s}}_t}, {{\bf{a}}_t}) \gets {\mathop{\mathbb{E}}}[\sum_{t = 0}^\infty  {{ \gamma ^t}}r({\bf{s}}_t, {\bf{a}}_t) \mid \pi_{\theta}]$.
        % \State Update $\theta_3$ with (\ref{pi3}).
        \State Train $\theta_3$ to maximize $Q^{\pi_{\theta_3}}$.
    \EndIf
\EndFor
\State \textbf{Output}: Optimal policies for $\pi^*_{\theta_3}$.
\end{algorithmic}
\end{algorithm}

\subsection{Communication Model}

The compressed data are transmitted to the edge server via wireless communications. Without loss of generality, we simplify the communication model by only considering the uplink latency and ignoring the latency for the downlink, since the communication resources spent on downlink including control commands and compress rate decision are often negligible. We also assume that the channel gain and transmission rate remain constant over a small time interval, and the two \glspl{ugv} have i.i.d. channel conditions. Here, we take one of the \gls{ugv} as an example, where the uplink transmission rate in the $t$-th time slot is given by
\begin{equation}
    r_{t}(t) = \text{B} \cdot \log_2 \left(1 + \frac{p(t) \cdot g(t)}{N_0}\right),
\end{equation}
where \( \text{B} \) denotes the bandwidth allocated to the \glspl{ugv}~\cite{shannon1948mathematical}. The signal-to-interference-plus-noise ratio (SINR) for the signals received at the edge server is expressed as 
\begin{equation}
    \text{SINR} = \frac{p(t) \cdot g(t)}{N_0},
\end{equation}
where \( p(t) \) is the transmission power of the \glspl{ugv}, and \( N_0 \) is the \gls{awgn}. The channel gain for the signals received at the edge server is 
\begin{equation}
    g(t) = d^{-\alpha} u_t,
\end{equation}
which accounts for the instantaneous perception of the communication environment. The distance \( d \) between the Jackal \glspl{ugv} and the edge server is the most significant factor affecting the channel gain. The path loss exponent is denoted by \( \alpha \), and \( u_t \) follows a Rayleigh distribution with a unit mean. 
% Generally, channel gain is a continuous variable. However, it can be quantified as \( g(t) \in \{-30 \ \text{dB}, \ -20 \ \text{dB}, \ -10 \ \text{dB}, \ 0 \ \text{dB}, \ 10 \ \text{dB}, \ 20 \ \text{dB}, \ 30 \ \text{dB}\} \), which simplifies model discrimination.
% Assuming the compressed normalized point cloud feature \( S_k \) has \( k \) components in float32, the number of bits required to be transmitted is given by:
% \begin{equation}
%     N^b = 32 \cdot k,
% \end{equation}
The uplink transmission latency is then given by
\begin{equation} \label{tup}
    T_{up} = \arg\min_{\substack{\int_0^{\delta} r(t) \, dt \geq N^b}} \delta \approx \frac{N^b}{r(t)},
\end{equation}
where \( N^b \) represents the total number of bits, the total number of bits \( N^b_f \) for front \gls{ugv} and \( N^b_r \) for rear \gls{ugv} are then given by
\begin{equation} \label{nb}
    N^b_f = n^{b} \cdot k_f(t), 
\end{equation}
\begin{equation} \label{nb} 
    N^b_r = n^{b} \cdot k_r(t),
\end{equation}
where $n^b$ is the number of bits in each component, $k_f(t)$ and $k_r(t)$  are the number of components in the point cloud features after the second-stage compression $ {\bf{z}}^f(t)$ and ${\bf{z}}^r(t)$. Given that the transmission rate $ r_{t}(t) $ in each time slot under a stationary channel is constant, the latency can be approximated as the ratio of the number of transmitted bits $N^b$ to $r_{t}(t)$.

\subsection{Edge Server}

The edge server first receives the compressed point cloud feature $ \hat{\bf{z}}^f(t)$ and $\hat{\bf{z}}^r(t)$ from the front \gls{ugv} and rear \gls{ugv}. With inversed \gls{lsa}, the compressed point cloud features will be decompressed with
\begin{align}
\label{decompress}
    \tilde{\bf{x}}(t) &= [\tilde{\bf{x}}^f(t), \tilde{\bf{x}}^r(t)]  \notag \\
    &= f_d^{-1}(\{{\hat{\bf{z}}}^f(t), {\hat{\bf{z}}}^r(t)\},\theta_d,\{k^f(t), k^r(t)\}).
\end{align}

In addition, we deploy three reinforcement learning agents on the edge server, which are responsible for the offload function 1) the front \gls{ugv} for navigation 2) the rear \gls{ugv} for collaborated conveyance, and 3) the compression ratio decision making, respectively. 
The two driving agents will infer driving commands for both \glspl{ugv} with observation $\tilde{\bf{x}}(t)$, where for the front \gls{ugv} the command is its angular speed $\theta^f(t)$ and for the rear \gls{ugv} is its angular and linear speeds $\theta^r(t)$ and $v^r(t)$. Then $\theta^f(t)$, $\theta^r(t)$, and $v^r(t)$ will be sent to the two \glspl{ugv} to be executed. More details about agent settings will be given in the next section.

% The decompressed feature $\bf{s}^f_t$ is then split into two features $\hat{\bf{x}}^f_t$ and $\hat{\bf{x}}^b_t$, where $\hat{\bf{x}}^f_t \in [0, 1]^{1 \times 128}$ and $\hat{\bf{x}}^b_t \in [0, 1]^{1 \times 128}$ are the compressed normalized point cloud data from the front ugv and the back ugv, which serve as observation for the two reinforcement learning agents. The agent calculate the action \(\mathbf{a}_{t}^{f}\) (the linear velocity \(v_f\) and angular velocity \(\theta_f\)) which maximizes the expected reward under \(\pi_{\theta}^{f}(a|s)\).  The other agent calculate the action \(\mathbf{a}_{t}^{b}\) (the linear velocity \(v_b\) and angular velocity \(\theta_b\)) which maximizes the expected reward under \(\pi_{\theta}^{b}(a|s)\).

\section{Problem Formulation}
% \Zhen{We have a use RL so we need algorithm section and  solution section. So you can clearly define what are the state, action n and rewards. Then you can directly define it as a DRL problems rather than MDP problems.}

% \xm{Problem formulation has been reworked}
%A:front jacakal parking

%We propose a two-phase \gls{drl} algorithm for different task-oriented components. The front ugv's navigation is governed by a path planning agent, $\pi_{\theta_1}^*$. The back ugv is controlled by agent $\pi_{\theta_2}^*$, which is responsible for following the front ugv while maintaining the parcel's designed trajectory. The compression agent $\pi_{\theta_3}^*$ determines the optimal compression ratio to adapt to different channel status. The setup of each agent will be detailed in the following sections.
We propose a two-phase \gls{drl} algorithm for different task-oriented components. In the first phase,
we train navigation agent $\pi_{\theta_1}$ and for the navigation of the front \gls{ugv} and collaborated conveyance agent $\pi_{\theta_2}$ for the rear \gls{ugv}, respectively. In the second phase, we train the dynamic compression agent $\pi_{\theta_3}$ with optimal $\pi_{\theta_1}^*$ and  $\pi_{\theta_2}^*$. $\pi_{\theta_3}$ determines the optimal task-oriented compression ratio for the dynamic compression algorithm.

% To optimize the performance of our framework, we propose to use a three-stage \gls{drl} algorithm for different task-oriented components. The front ugv is navigated by a path planning agent $\pi_{\theta}^1$. The back ugv is controlled by $\pi_{\theta}^2$ to follow the front ugv while keeping the parcel on trajectory. The compression ratio will be given by the compression agent $\pi_{\theta}^{3*}$ to adapt to the current channel status. 
% Two \gls{drl} agents will give control commands to navigate the front and back ugv, and a third \gls{drl} agent will provide optimal compression ratios in every time slot. Specifically, we propose to use the \gls{ppo} algorithm in all \gls{drl} agents for its simplicity and sampling efficiency.
\subsection{Navigation}
The navigation agent with \gls{drl} policy $\pi_{\theta_1}$ navigates the front path and produces $\theta^f(t)$ and ensures that it reaches the given target position. 
\subsubsection{State}
In the $t$-th time slot, the state of $\pi_{\theta_1}$ agent is
\begin{equation}
\begin{aligned}
    {\bf{s}}_{t}^{[1]} = \tilde{\bf{x}}(t),
\end{aligned}
\end{equation}
where $\tilde{\bf{x}}(t) \in [0, 1]^{1 \times n}$ is the normalized decompressed point cloud data at the edge server with $1 \times n$ dimensions.
\subsubsection{Action}
The action in the $t$-th time slot of $\pi_{\theta_1}$ agent is the normalized angular velocity of the front \gls{ugv}
\begin{align}
    {\bf{a}}_{t}^{[1]} = \theta^f(t),
\end{align}
where $\theta^f(t) \in [0, 1]$ is the angular speed command for the front \gls{ugv}.
\subsubsection{Reward}
The reward of $\pi_{\theta_1}$ agent is given by
\begin{equation}
\begin{aligned}
\label{r1}
    {r}^{[1]}({\bf{s}}_{t}^{[1]}, {\bf{a}}_{t}^{[1]}) = w_1 \cdot &\text{MSE}\left({{\bf{p}}^f(t)}, {{\bf{p}}_{tar}^f}\right)  \\  +&w_2 \cdot |\phi^f(t) - \phi^f_{tar}|,
\end{aligned}
\end{equation}
where ${{\bf{p}}^f(t)} = [x^f(t), y^f(t)]$ is the position of the front \gls{ugv} in the $t$-th timeslot, ${{\bf{p}}_{tar}^f} = [x^f_{tar}, y^f_{tar}]$ is its target position; $\phi^f(t)$ and $\phi^f_{tar}$ are the orientation of the front \gls{ugv} in the $t$-th timeslot and the target orientation. The $\text{MSE}(\cdot, \cdot)$ operator denotes the \gls{mse} of two positions, defined as
\begin{equation}
\begin{aligned}
    \text{MSE}({{\bf{p}}_1}, {{\bf{p}}_2}) = \sqrt{(x_1-x_2)^2 + (y_1-y_2)^2}.
\end{aligned}
\end{equation}
\subsection{Cooperative Conveyance}
   The cooperative conveyance agent with \gls{drl} policy $\pi_{\theta_2}$ controls the rear \gls{ugv} and ensures that it follows the front \gls{ugv} and gives adequate force to the parcel so that it is held between the two Jackals in the same trajectory. 
\subsubsection{State}
In the $t$-th time slot, the state of $\pi_{\theta_2}$ agent is
\begin{equation}
\begin{aligned}
    {\bf{s}}_{t}^{[2]} = \tilde{\bf{x}}(t).
\end{aligned}
\end{equation}
\subsubsection{Action}
The action in the $t$-th time slot of $\pi_{\theta_2}$ agent is the normalized angular and linear velocity of the rear \gls{ugv}
\begin{equation}
\begin{aligned}
    {\bf{a}}_{t}^{[2]} = [\theta^{r}(t), v^{r}(t)],
\end{aligned}
\end{equation}
where $\theta^{r}(t) \in [0, 1]$ and $v^{r}(t) \in [0, 1]$ are the normalized angular and linear velocities for the rear \gls{ugv}, respectively.
\subsubsection{Reward}
The reward of $\pi_{\theta_2}$ agent is given by
\begin{equation}
\begin{aligned}
\label{r2}
    {r}^{[2]}(&{\bf{s}}_{t}^{[2]}, {\bf{a}}_{t}^{[2]}) = w_3 \cdot \text{MSE}({{\bf{p}}^f(t)}, {{\bf{p}}^r(t)}) \\
    &+ w_4 \cdot \text{MSE}({{\bf{p}}^c(t)}, {{\bf{p}}^r(t)}) + w_5 \cdot p_{m},
\end{aligned}
\end{equation}
where ${{\bf{p}}^r(t)} = [x^r(t), y^r(t)]$ is the position of the rear \gls{ugv} in the $t$-th timeslot, ${{\bf{p}}^c(t)} = [x^p(t), y^p(t)]$ is the position of the cargo box in the $t$-th timeslot, and $p_{m}$ is a penalty given when the orientations of the two \glspl{ugv} are misaligned over a threshold $\phi_T$, denoted as
\begin{equation}
    \begin{aligned}
    \label{poisson}
    p_{m} = 
    & \begin{cases}
    \;\;0\;, \;\;\; \text{when} \; |\phi^f(t) - \phi^r(t)| < \phi_T, \\
    -1, \;\;\; \text{otherwise}. \;
    \end{cases}
     \end{aligned}
\end{equation}
\subsection{Dynamic Compression}
In the $t$-th time slot, the compression ratio $\rho^f$ and $\rho^r$ is decided by the dynamic compression agent with \gls{drl} policy $\pi_{\theta_3}$.
\subsubsection{State}
In the $t$-th time slot, the state of $\pi_{\theta_3}$ agent is the merged decompressed point cloud data with the channel gains:
\begin{equation}
\begin{aligned}
    {\bf{s}}^{[3]}_t = [\tilde{\bf{x}}(t), g^f(t), g^r(t)],
\end{aligned}
\end{equation}
where $\tilde{\bf{x}}(t) \in [0, 1]^{1 \times n}$ is the merged decompressed point cloud data with $1 \times n$ dimensions, $g^f(t)$ and $g^r(t)$ are the channel gains of front and rear \glspl{ugv}, respectively.
\subsubsection{Action}
The action in the $t$-th time slot of $\pi_{\theta_3}$ agent is the compression parameters for both dynamic compressors, denoted by 
\begin{equation}
\begin{aligned}
    {\bf{a}}_{t}^{[3]} = [k^f(t), k^r(t)],
\end{aligned}
\end{equation}
where $k^f(t)$ and $k^r(t)$ are the compression parameters for the front and rear \glspl{ugv}, respectively.
\subsubsection{Reward}
The reward of $\pi_{\theta_3}$ agent is a weighted sum of the distance between the cargo box and the loading area and the communication cost, denoted by
\begin{equation}
\begin{aligned}
\label{r3}
    {r}^{[3]}({\bf{s}}_{t}^{[3]}, {\bf{a}}_{t}^{[3]}) = w_6 \cdot &\text{MSE}({{\bf{p}}^c(t)}, {{\bf{p}}_{tar}^c}) \\ - &w_7 \cdot \text{max}(T^f_{up}, T^r_{up}),
\end{aligned}
\end{equation}
where ${{\bf{p}}_{tar}^c}$ is the target position of the cargo box and $T_{up}$ is the uplink transmission latency in the $t$-th time slot.

\subsection{Solution}
The proposed two-phase \gls{drl} problem is shown in Algorithm \ref{algorithm1} based on the \gls{ppo} algorithm~\cite{schulman2017proximal}. We pre-train two \textit{PointNet++} neural networks by embedding individual DRL in the experiment environment without the integration of three agents. 

In the phase 1, after the feature vector $\tilde{\bf{x}}_t$ is transmitted to the edge server, the $\pi_{\theta_1}$ and $\pi_{\theta_2}$ generate ${\bf{a}}^{[1]}_t$ with the observation $\tilde{\bf{x}}_t$ simultaneously, and ${\bf{a}}^{[2]}_t$ transmit them back to \glspl{ugv}. Then, by executing ${\bf{a}}^{[1]}_t$ and ${\bf{a}}^{[2]}_t$, the instantaneous rewards $r^{[1]}({\bf{s}}_t^{[1]}, {\bf{a}}_t^{[1]})$ and $r^{[2]}({\bf{s}}_t^{[2]}, {\bf{a}}_t^{[2]})$ are also obtained. Thus, by sampling the batch from the collected tuple $\{\tilde{\bf{x}}_t, {\bf{a}}^{[1]}_t, {\bf{a}}^{[2]}_t, r^{[1]}({\bf{s}}_t^{[1]}, {\bf{a}}_t^{[1]}), r^{[2]}({\bf{s}}_t^{[2]}, {\bf{a}}_t^{[2]})\}$ and train
$\theta_1$ and $\theta_2$, $\pi^*_{\theta_1}$ and $\pi^*_{\theta_2}$ are obtained. In the phase 2, with the optimal and fixed $\pi^*_{\theta_1}$ and $\pi^*_{\theta_2}$ obtained in the phase 1, $\pi_{\theta_3}$ is optimized by sampling the batch from the collected tuple $\{\tilde{\bf{x}}_t, g^f(t), g^r(t), {\bf{a}}^{[3]}_t, r^{[3]}({\bf{s}}_t^{[3]}, {\bf{a}}_t^{[3]})\}$.

\begin{figure}[h]
    \centering
    \includegraphics[width=\linewidth,angle = 0]{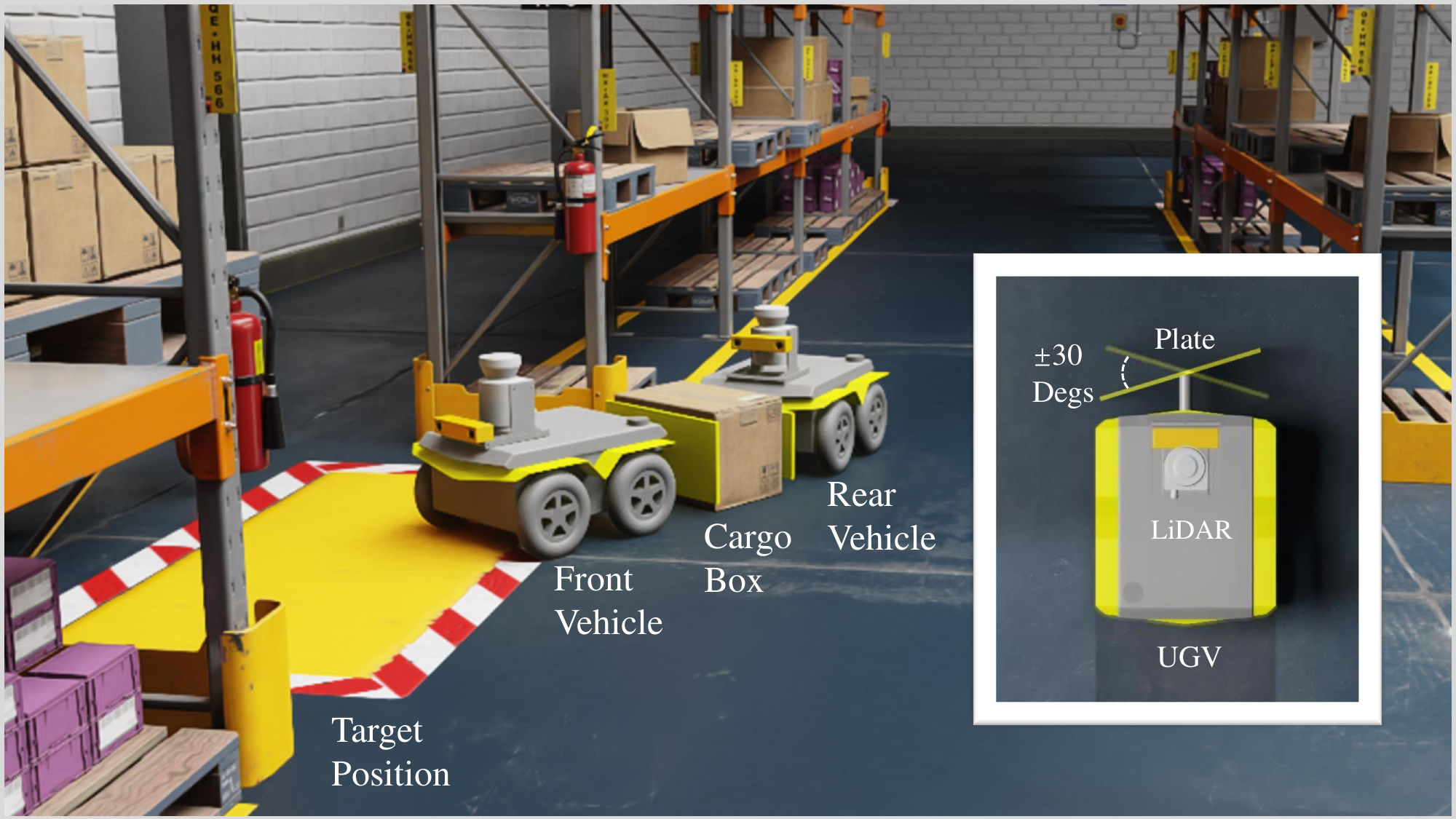}
    \caption{Prototype design in Isaac Sim (The demonstration video is available at\url{https://youtu.be/egEIaBiVmpA}).}.
    \label{fig:warehouse}
\end{figure}

\section{Experimental Setup}

\begin{figure*}
  %\centering
  \subfigure[Average reward per step in each training episode for front controlagent $\pi_{\theta_1}$.]{%caption of the subfigure
  \label{fig: reward_ugv_1}%%label for first subfigure
  \includegraphics[scale=0.28]{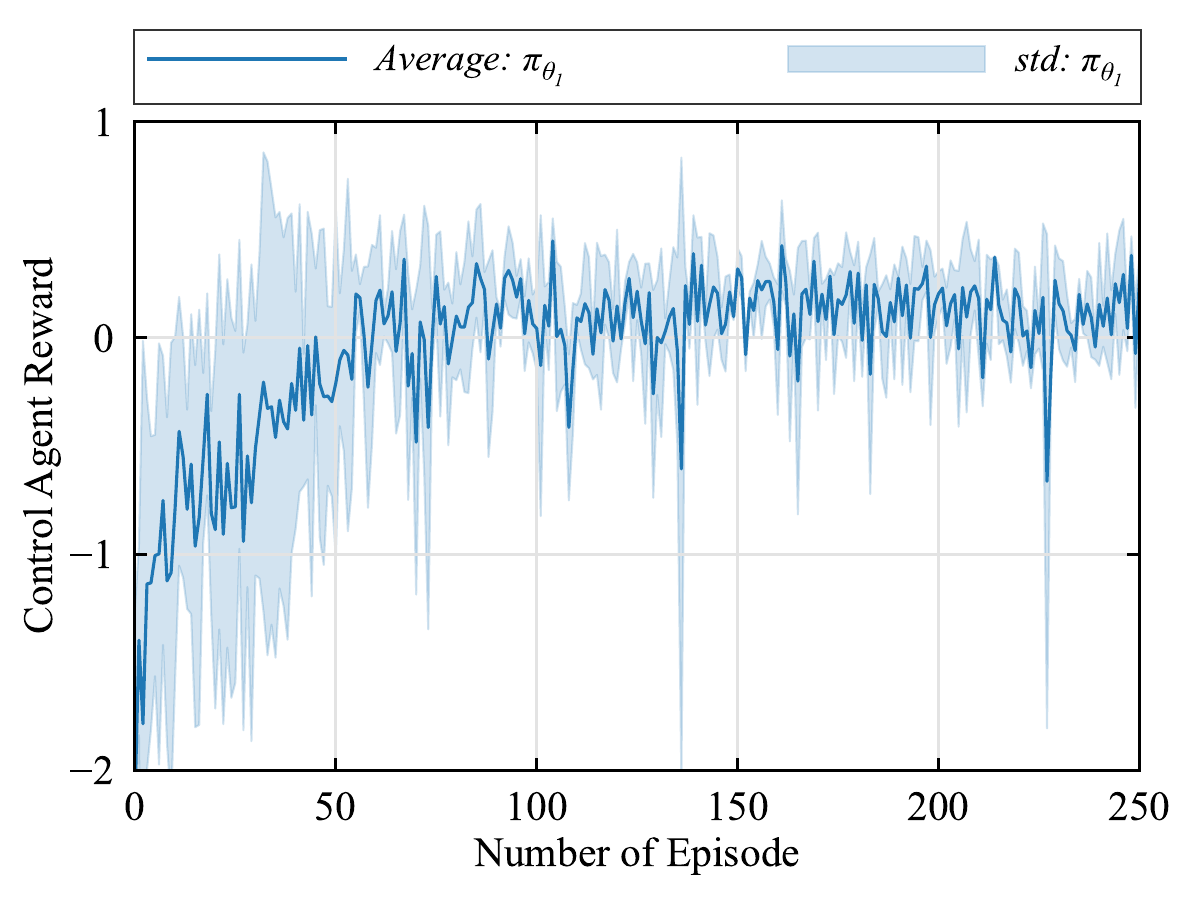}
  }
  \subfigure[Average reward per step in each training episode for rear control agent $\pi_{\theta_2}$.]{%caption of the subfigure
  \label{fig: reward_ugv_2}%%label for second subfigure
  \includegraphics[scale=0.28]{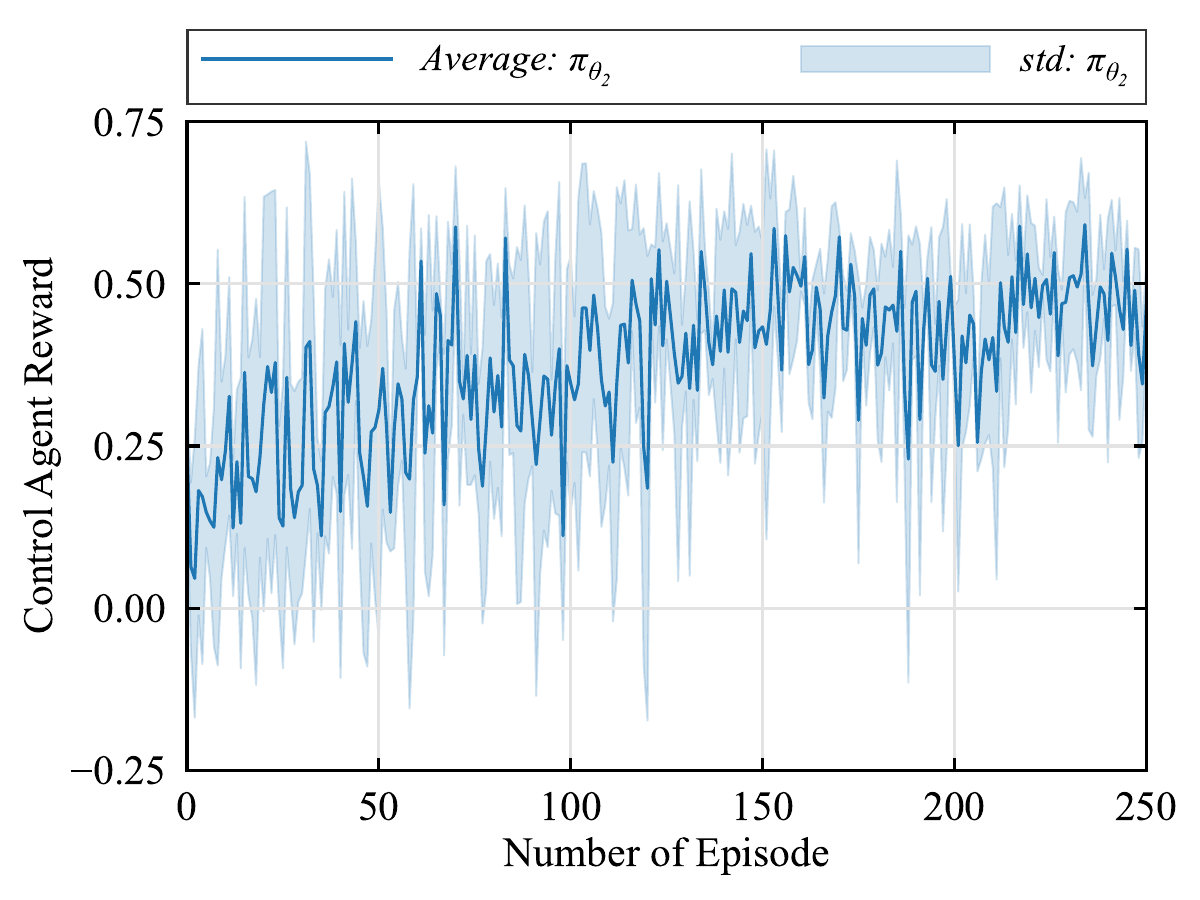}}
  \subfigure[Average reward per step in each training episode for compression agent $\pi_{\theta_3}$.]{%caption of the subfigure
  \label{fig: reward_compression}%%label for second subfigure
  \includegraphics[scale=0.28]{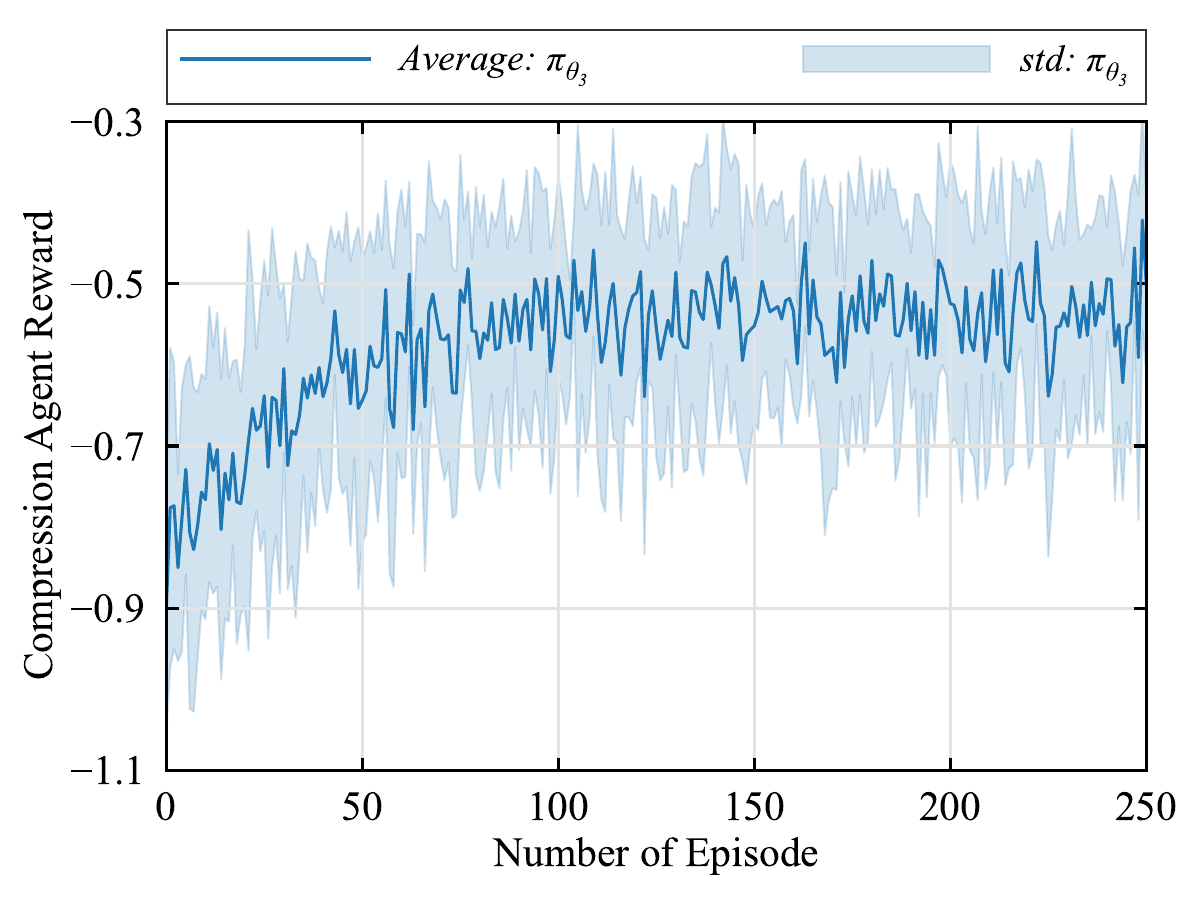}}  
  \caption{Performance evaluation for three \gls{drl} agent $\pi_{\theta_1}$, $\pi_{\theta_2}$ and $\pi_{\theta_3}$.} % in the training stage}
\end{figure*}

To evaluate the performance of the proposed framework, we establish our simulation platform on Nvidia Isaac Sim~\cite{isaac_sim} for its realistic physical simulator and integral interface for robotic systems. As shown in Fig.~\ref{fig:warehouse}, two Clearpath Jackal \glspl{ugv}~\cite{jackal} are modeled in a warehouse environment. Notably, for each Jackal, a plate is mounted on the front/rear side with a hinge to allow passive rotations at $\pm 30$~degs. The plates help the Jackal \glspl{ugv} curve without releasing the carried cargo box. The two Jackal \glspl{ugv} with their attached plates together make our cooperative conveyance system~\cite{8927467} for cargo boxes in the proposed framework. The sensory system is built by two LiDARs mounted on the top of each Jackal \gls{ugv}. Point cloud data will be sampled by each LiDA at a certain frequency.. 
The task in this experiment is to move the cargo box with our conveyance system to the target position, which is labeled as the yellow zone in Fig.~\ref{fig:warehouse}. In each time slot, the channel gains $g^f(t)$ and $g^r(t)$ are sampled from an average distribution, with values taken from ${\{-30 \text{dB}, \ -20 \text{dB}, \ -10 \text{dB}, \ 0 \text{dB}, \ 10 \text{dB}, \ 20 \text{dB}, \ 30 \text{dB}\}}$. The dynamic channel compression parameter $k^f(t)$ and $k^r(t)$ takes value from $\{1, 2, 3,...127\}$. Thus, the compressed normalized point cloud feature has $k^f(t)$ and $k^r(t)$ components in float32 where each float32 data type consists of 32 bits.

% For the purposes of this research, a detailed simulation of a warehouse environment was constructed within Isaac Sim, as depicted in Fig.~\ref{fig:warehouse}. This simulation meticulously replicates a typical storage facility, equipped with a variety of features to enhance its realism. Two Jackal UGVs were modeled, to suit the specific requirements of the task, modifications were made to the original Jackal design. Notably, a board was mounted on the front of one Jackal and on the back of the other, as it allows the two UGVs to clamp and transport parcels by aligning the board between them. Moreover, the RtxLidar was mounted on both two UGVs, which are only input devices in our framework.

% The extensive library of textures and models available in Isaac Sim was employed to enrich the environment's authenticity. This included a selection of materials for different surfaces, such as floor tiles and walls, as well as models of common warehouse items like shelves and boxes. These elements were strategically arranged to create the cluttered yet structured layout characteristic of a conventional warehouse, providing a realistic setting for testing the navigational and obstacle avoidance capabilities of the Jackal UGVs.

% Based on the overall task, the front UGV and the rear UGV have their own specific tasks. The front UGV is responsible for 1) navigating into the designated parking and unloading area, and 2) avoiding obstacles such as shelves and trolleys during movement. The rear UGV is responsible for 1) following the front UGV to push cargo, and 2) avoiding obstacles.

\section{Simulation Result}

We evaluate 1) the performance of \gls{drl} training process for $\pi_{\theta_1}$, $\pi_{\theta_2}$ and $\pi_{\theta_3}$. 2) the task completion rate for the cooperative conveyance system with the integration of $\pi_{\theta_3}$. 3) the effectiveness of the proposed $\pi_{\theta_3}$.

\subsection{Evaluation of Task Completion}
As shown in Fig.~\ref{fig: reward_ugv_1} and Fig.~\ref{fig: reward_ugv_2}, to evaluate the performance of task completion for two Jackals \glspl{ugv} without the compression \gls{drl} module, we simultaneously train the dual \gls{ppo} algorithms for two Jackals \glspl{ugv} control agents, $\pi_{\theta_1}$ and $\pi_{\theta_2}$, over $250$ episodes for $3$ times.
$\pi_{\theta_1}$ converged after 80 training epochs, while $\pi_{\theta_2}$
converged after 140 training epochs. 

% Once the cooperative conveyance system is trained, it can consistently complete the conveyance tasks, which will be discussed in the following sub-section.
% %  \begin{figure}
% \centering
% \includegraphics[scale=0.4]{Results/Reward_U.pdf}
% \caption{Average reward per step in each training episode for agent $\pi_{\theta_1}$ and $\pi_{\theta_2}$.}
% \label{fig: reward_ugvs}
% \end{figure}

\begin{figure}
\centering
\includegraphics[scale=0.37]{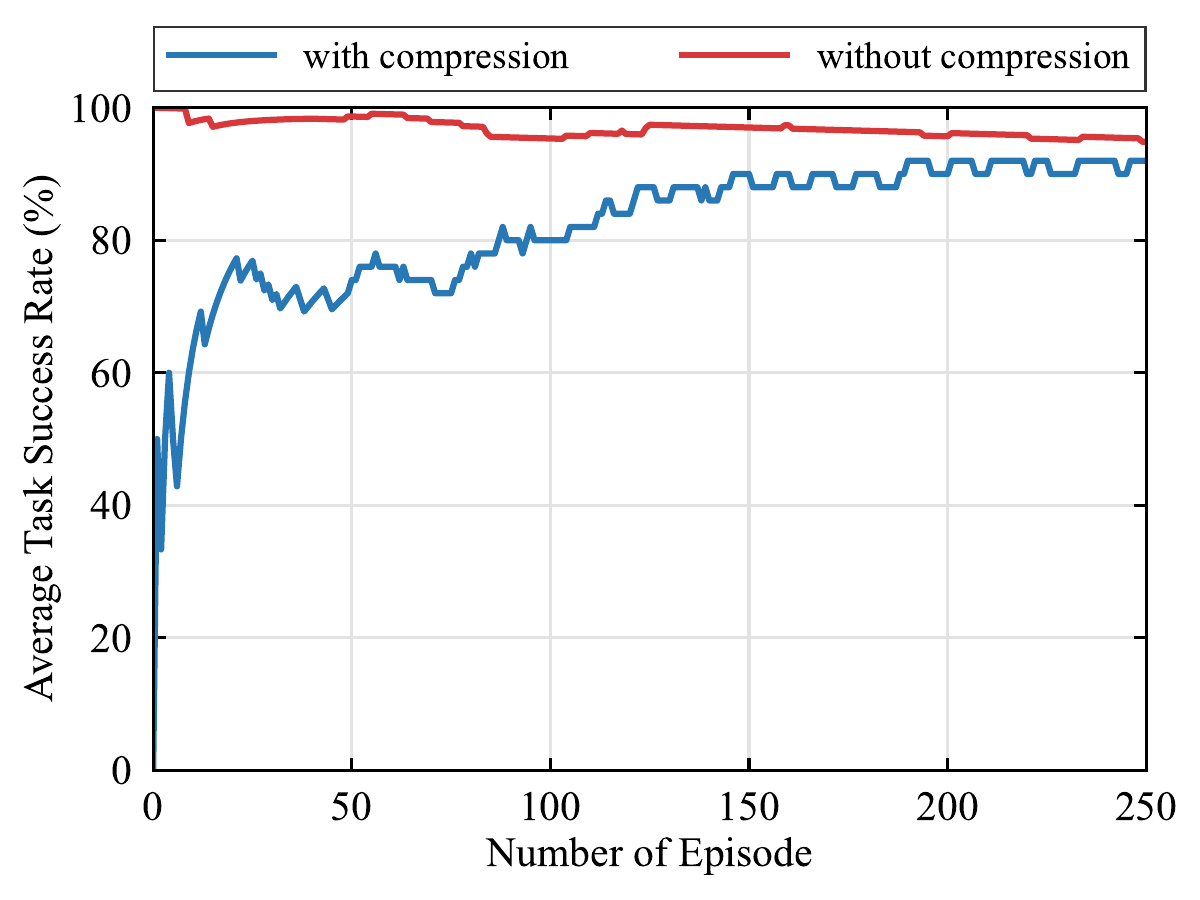}
\caption{Average task success rate in each training episode.}
\label{fig: task}
\end{figure}

\subsection{Evaluation of Compression Agent}
\begin{figure}
\centering
\includegraphics[scale=0.37]{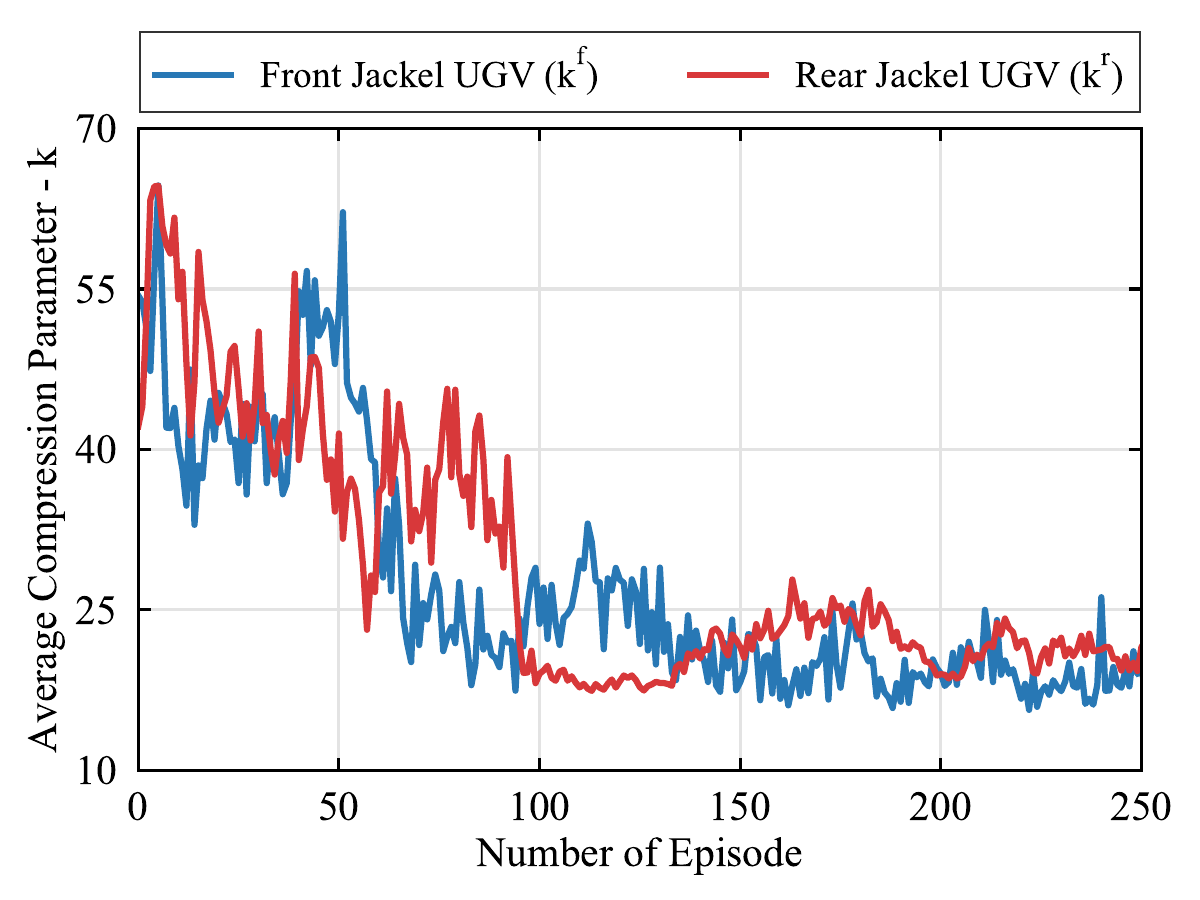}
\caption{Average compression Parameter in each training episode.}
\label{fig: action_compression}
\end{figure}

Fig.~\ref{fig: reward_compression} illustrates the changes in the average reward and its standard deviation for $\pi_{\theta_3}$. The average reward value generally shows an upward trend, and the fluctuations of the compression ratio gradually decrease. Additionally, Fig.~\ref{fig: action_compression} displays the changes in the compression parameters over $250$ training episodes. Both compression ratios $\rho^f$ and $\rho^r$ decrease during the training process and converge at $k^f = 19, k^r = 22$ after $185$ training episodes, which means that the overall compression ratio decreases to approximately $0.32 \%$ compared to the raw point cloud data, which initially consisted of approximately $20,000$ points before compression.

\subsection{Evaluation of Task completion performance under Dynamic Compression}
As shown in Fig.~\ref{fig: task}, we evaluate the task completion rate after incorporating the compression agent. Specifically, we define the average task success rate as the number of successful attempts out of 100 repetitions of the experiment in the current training episode. The weights for $\pi_{\theta_3}$, are set to $\omega_6 = 0.8$ and $\omega_7 =0.2$. The reward of both two control agents for two Jackals \glspl{ugv} converges after training of $130$ episodes. Our proposed framework successfully completes the task with an approximate average task success rate of $90$\% after convergence, which is close to the success rate without the compression model.

% \subsection{Evaluation of Different Channel \gls{snr}}
% We evaluate the impact of different channel \glspl{snr} on task completion performance. Specifically, we consider the performance metrics: success rate and communication reliability under varying \gls{snr} conditions. Tab.~\ref{} presents the results across different \gls{snr} levels. Our analysis demonstrates the relationship between channel \gls{snr} and task completion performance. Higher \gls{snr} values generally lead to better success rates and higher communication reliability, while lower \gls{snr} values negatively impact these performance metrics. This analysis provides valuable insights into optimizing system performance under varying channel conditions.

% \begin{figure}
% \centering
% \includegraphics[scale=0.4]{Results/Reward_C.pdf}
% \caption{Average reward per step in each training episode for agent $\pi_{\theta_3}$.}
% \label{fig: reward_compression}
% \end{figure}

\subsection{Evaluation of Different Weights}

We further evaluate the robustness of the proposed framework by adjusting different weights $\omega_6$ and $\omega_7$. Tab.~\ref{teble_weight} shows the results with varying weights $\omega_6$ and $\omega_7$. As $\omega_6$ increases from $\omega_6 = 0.5$ to $0.99$ and $\omega_7$ decreases from $\omega_7 = 0.5$ to $0.01$, the task success rate increases significantly to 96\%. However, the average compression parameter $k^{f}(t)$ and $k^{r}(t)$ will increase with, occupying more communication load.

\begin{table}
\centering
\caption{Evaluation of Different Weight}
\label{table_weight}
\resizebox{0.95\linewidth}{!}{
  \begin{tabular}{|c|c|c|c|c|c|c|}
    \hline
    \rowcolor[RGB]{230,230,230}
    $\omega_6$ & 0.99 & 0.9 & 0.8 & 0.7 & 0.6 & 0.5 \\
    \hline
    \rowcolor[RGB]{230,230,230}
    $\omega_7$ & 0.01 & 0.1 & 0.2 & 0.3 & 0.4 & 0.5 \\
    \hline
    success rate (\%) & 96 & 93 & 91 & 82 & 69 & 53 \\
    \hline
    average $k^f(t)$ & 103 & 43 & 19 & 14 & 7 & 6 \\
    \hline
    average $k^r(t)$ & 110 & 49 & 22 & 11 & 9 & 5 \\
    \hline
  \end{tabular}
}
\end{table}

% \subsection{Jackal UGVs transportation task}
 
% Fig.~\ref{fig:EVRMSE} shows that the EVR gradually increases and stabilizes as the number of components increases. The EVR indicates how much of the variance in the original data is captured by the compressed data. Higher EVR values suggest better retention of the original data's variance. It also illustrates that the MSE gradually decreases and stabilizes, which measures the average squared difference between the original and reconstructed data. Lower MSE values indicate higher reconstruction accuracy.

% These initial results indicate that while the compression \gls{drl} module reduces the dimensionality of the feature vector significantly, it also introduces a decrease in the task success rate. However, this compression is crucial for efficient communication and storage, particularly in resource-constrained environments. Further optimization of the \gls{drl} module and the feature inversion process is expected to improve both the compression efficiency and the overall system performance.

\section{Conclusion}
We presented a task-oriented edge-assisted cooperative data compression, communication, and computing framework for \gls{ugv}-enhanced warehouse logistics. The warehouse logistics collaboration task, where two \glspl{ugv} cooperatively carry a cargo box to the target position, was successfully offloaded to the edge server. In addition, a two-stage \gls{drl}-based point cloud data compression algorithm was deployed on the edge server and the \glspl{ugv}, where the dynamic compression ratio of two \glspl{ugv} $\rho^f$ and $\rho^r$ was dynamically and collaboratively determined based on the task requirements. The experimental result showed that the system can complete the task with a success rate of 
% 80\% without the compression DRL module and 
$90\%$ with the compression module ($\omega_6=0.8$, $\omega_7=0.2$), with an average compression ratio of $ 3.2\%$ after the static compression and an average compression ratio of $0.3\%$ after the dynamic compression.

% \input{1_intro}
% \input{2_related_work}
% \input{3_system_model}
% \input{4_result}
% \input{5_conclusion}

% If use 'cite' package
%\bibliographystyle{unsrt}
%\bibliography{reference}

\printbibliography

@String{ICASSP = "Proc. IEEE ICASSP"}

@article{nvidia_warehouse,
  title={AI-Enabled Warehouse Logistics},
  author={NVIDIA},
  journal={NVIDIA},
  year={2024},
  note={\url{https://www.nvidia.com/en-gb/industries/retail/warehouse-logistics/}}
}

@article{Xie2021DeepSC,
  title={DeepSC: Deep Learning Enabled Semantic Comm. Systems},
  author={Xie, Huiqiang and Qin, Zhiwei and Tao, Xiaoming and Lu, Guanding and Wu, Di},
  journal={IEEE Trans. Wireless Commun.},
  volume={20},
  number={8},
  pages={5120--5133},
  year={2021},
  publisher={IEEE}
}

@article{Zhang2024Optimization,
  title={Optimization of Image Transmission in Cooperative Semantic Comm. Networks},
  author={Zhang, Wenjing and Wang, Yining and Chen, Mingzhe and Luo, Tao and Niyato, Dusit},
  journal={IEEE Trans. Wireless Commun.},
  volume={23},
  number={2},
  pages={861--872},
  year={2024},
  publisher={IEEE}
}

@ARTICLE{Talli2023Dynamic,
  author={Talli, Pietro and Pase, Francesco and Chiariotti, Federico and Zanella, Andrea and Zorzi, Michele},
  journal={IEEE Trans. Commun.}, 
  title={Effective Comm. with Dynamic Feature Compression}, 
  year={2024},
  volume={},
  number={},
  pages={1--1},
  doi={10.1109/TCOMM.2024.3385922}
}

@ARTICLE{9910575,
  author={Qiao, Yue and Fu, Yusun and Yuan, Muyun},
  journal={IEEE Internet Things J.}, 
  title={Comm.–Control Co-Design in Wireless Networks: A Cloud Control AGV Example}, 
  year={2023},
  volume={10},
  number={3},
  pages={2346--2359},
  doi={10.1109/JIOT.2022.3211766}
}

@ARTICLE{yufeng,
    author={Diao, Yufeng and Meng, Zhen and Xu, Xiangmin and She, Changyang and Zhao, Philip G.},
    journal={IEEE INFOCOM 2024-IEEE Conf. Comput. Commun. Workshops (INFOCOM WKSHPS)},
    title={Task-Oriented Source-Channel Coding Enabled Autonomous Driving Based on Edge Computing},
    year={2024},
    month={2}
}

@ARTICLE{9509294,
  author={Nguyen, Dinh C. and Ding, Ming and Pathirana, Pubudu N. and Seneviratne, Aruna and Li, Jun and Niyato, Dusit and Dobre, Octavia and Poor, H. Vincent},
  journal={IEEE Internet Things J.}, 
  title={6G Internet of Things: A Comprehensive Survey}, 
  year={2022},
  volume={9},
  number={1},
  pages={359--383},
  doi={10.1109/JIOT.2021.3103320}
}

@ARTICLE{8493155,
  author={Chen, Xianfu and Zhang, Honggang and Wu, Celimuge and Mao, Shiwen and Ji, Yusheng and Bennis, Medhi},
  journal={IEEE Internet Things J.}, 
  title={Optimized Computation Offloading Performance in Virtual Edge Computing Systems Via Deep Reinforcement Learning}, 
  year={2019},
  volume={6},
  number={3},
  pages={4005--4018},
  doi={10.1109/JIOT.2018.2876279}
}

@ARTICLE{8016573,
  author={Mao, Yuyi and You, Changsheng and Zhang, Jun and Huang, Kaibin and Letaief, Khaled B.},
  journal={IEEE Commun. Surv. Tutor.}, 
  title={A Survey on Mobile Edge Computing: The Comm. Perspective}, 
  year={2017},
  volume={19},
  number={4},
  pages={2322--2358},
  doi={10.1109/COMST.2017.2745201}
}

@ARTICLE{10554663,
  author={Chaccour, Christina and Saad, Walid and Debbah, Mérouane and Han, Zhu and Poor, H. Vincent},
  journal={IEEE Commun. Surv. Tutor.}, 
  title={Less Data, More Knowledge: Building Next Generation Semantic Comm. Networks}, 
  year={2024},
  volume={},
  number={},
  pages={1--1},
  doi={10.1109/COMST.2024.3412852}
}

@article{schulman2017proximal,
  title={Proximal policy optimization algorithms},
  author={Schulman, John and Wolski, Filip and Dhariwal, Prafulla and Radford, Alec and Klimov, Oleg},
  journal={arXiv preprint arXiv:1707.06347},
  year={2017}
}

@article{dinh2013survey,
  title={A Survey of Mobile Cloud Computing: Architecture, Applications, and Approaches},
  author={Dinh, Hoang T and Lee, Chonho and Niyato, Dusit and Wang, Ping},
  journal={Wirel. Commun. Mob. Comput.},
  volume={13},
  number={18},
  pages={1587--1611},
  year={2013},
  publisher={Wiley Online Library}
}

@ARTICLE{10422886,
  author={Meng, Zhen and She, Changyang and Zhao, Guodong and Imran, Muhammad A. and Dohler, Mischa and Li, Yonghui and Vucetic, Branka},
  journal={IEEE Wirel. Commun.}, 
  title={Task-Oriented Metaverse Design in the 6G Era}, 
  year={2024},
  volume={31},
  number={3},
  pages={212--218},
  doi={10.1109/MWC.019.2200605}
}

@ARTICLE{9955525,
  author={Gündüz, Deniz and Qin, Zhijin and Aguerri, Inaki Estella and Dhillon, Harpreet S. and Yang, Zhaohui and Yener, Aylin and Wong, Kai Kit and Chae, Chan-Byoung},
  journal={IEEE J. Sel. Areas Commun.}, 
  title={Beyond Transmitting Bits: Context, Semantics, and Task-Oriented Comm.}, 
  year={2023},
  volume={41},
  number={1},
  pages={5--41},
  doi={10.1109/JSAC.2022.3223408}
}

@ARTICLE{10370739,
  author={Meng, Zhen and Chen, Kan and Diao, Yufeng and She, Changyang and Zhao, Guodong and Imran, Muhammad Ali and Vucetic, Branka},
  journal={IEEE J. Sel. Areas Commun.}, 
  title={Task-Oriented Cross-System Design for Timely and Accurate Modeling in the Metaverse}, 
  year={2024},
  volume={42},
  number={3},
  pages={752--766},
  doi={10.1109/JSAC.2023.3345398}
}

@article{3GPP,
  title         = "\emph{Study on scenarios and requirements for next generation access technologies}",
  journal      = {\rm{document 3GPP, TSG RAN TR38.913 R14}},
  month         = Jun,
  year          = "2017",
}

@book{richards2017warehouse,
  title={Warehouse Management: A Complete Guide to Improving Efficiency and Minimizing Costs in the Modern Warehouse},
  author={Richards, Gwynne},
  year={2017},
  publisher={Kogan Page Publishers}
}

@misc{he2024qoemaximizationmultipleuavassistedmultiaccess,
  title={QoE Maximization for Multiple-UAV-Assisted Multi-Access Edge Computing: An Online Joint Optimization Approach}, 
  author={Long He and Geng Sun and Zemin Sun and Qingqing Wu and Jiawen Kang and Dusit Niyato and Zhu Han and Victor C. M. Leung},
  year={2024},
  eprint={2406.11918},
  archivePrefix={arXiv},
  primaryClass={eess.SY},
  url={https://arxiv.org/abs/2406.11918}, 
}

@ARTICLE{10007839,
  author={Liu, Shumei and Yu, Yao and Lian, Xiao and Feng, Yuze and She, Changyang and Yeoh, Phee Lep and Guo, Lei and Vucetic, Branka and Li, Yonghui},
  journal={IEEE J. Sel. Areas Commun.}, 
  title={Dependent Task Scheduling and Offloading for Minimizing Deadline Violation Ratio in Mobile Edge Computing Networks}, 
  year={2023},
  volume={41},
  number={2},
  pages={538--554},
  doi={10.1109/JSAC.2022.3233532}
}

@ARTICLE{9406800,
  author={Penmetcha, Manoj and Min, Byung-Cheol},
  journal={IEEE Access}, 
  title={A Deep Reinforcement Learning-Based Dynamic Computational Offloading Method for Cloud Robotics}, 
  year={2021},
  volume={9},
  number={},
  pages={60265--60279},
  doi={10.1109/ACCESS.2021.3073902}
}

@INPROCEEDINGS{8927467,
  author={Kumagai, Taichi and Yasuda, Shinya and Yoshida, Hiroshi},
  booktitle={IECON 2019 - 45th Annual Conf. IEEE Ind. Electron. Soc.}, 
  title={A Prototype of a Cooperative Conveyance System by Wireless-Network Control of Multiple Robots}, 
  year={2019},
  volume={1},
  number={},
  pages={231--236},
  doi={10.1109/IECON.2019.8927467}
}

@INPROCEEDINGS{9414831,
  author={Shao, Jiawei and Zhang, Haowei and Mao, Yuyi and Zhang, Jun},
  booktitle={ICASSP 2021 - 2021 IEEE Int. Conf. Acoust., Speech Signal Process. (ICASSP)}, 
  title={Branchy-GNN: A Device-Edge Co-Inference Framework for Efficient Point Cloud Processing}, 
  year={2021},
  volume={},
  number={},
  pages={8488--8492},
  doi={10.1109/ICASSP39728.2021.9414831}
}

@ARTICLE{10183796,
  author={He, Guojun and Feng, Mingjie and Zhang, Yu and Liu, Guanghua and Dai, Yueyue and Jiang, Tao},
  journal={IEEE Wirel. Commun.}, 
  title={Deep Reinforcement Learning Based Task-Oriented Communication in Multi-Agent Systems}, 
  year={2023},
  volume={30},
  number={3},
  pages={112--119},
  doi={10.1109/MWC.003.2200469}
}

@ARTICLE{10164147,
  author={Xu, Yujie and Zhou, Hui and Deng, Yansha},
  journal={IEEE Commun. Lett.}, 
  title={Task-Oriented Semantics-Aware Communication for Wireless UAV Control and Command Transmission}, 
  year={2023},
  volume={27},
  number={8},
  pages={2232--2236},
  doi={10.1109/LCOMM.2023.3290109}
}

@ARTICLE{10570351,
  author={Mostaani, Arsham and Vu, Thang X. and Habibi, Hamed and Chatzinotas, Symeon and Ottersten, Björn},
  journal={IEEE Trans. Commun.}, 
  title={Task-Oriented Communication Design at Scale}, 
  year={2024},
  volume={},
  number={},
  pages={1--1},
  doi={10.1109/TCOMM.2024.3416898}
}

@INPROCEEDINGS{10520522,
  author={Sagduyu, Yalin E. and Erpek, Tugba and Yener, Aylin and Ulukus, Sennur},
  booktitle={2023 IEEE Future Netw. World Forum (FNWF)}, 
  title={Multi-Receiver Task-Oriented Communications via Multi-Task Deep Learning}, 
  year={2023},
  volume={},
  number={},
  pages={1--6},
  doi={10.1109/FNWF58287.2023.10520522}
}

@ARTICLE{9837474,
  author={Shao, Jiawei and Mao, Yuyi and Zhang, Jun},
  journal={IEEE Trans. Wirel. Commun.}, 
  title={Task-Oriented Communication for Multidevice Cooperative Edge Inference}, 
  year={2023},
  volume={22},
  number={1},
  pages={73--87},
  doi={10.1109/TWC.2022.3191118}
}

@ARTICLE{9606667,
  author={Shao, Jiawei and Mao, Yuyi and Zhang, Jun},
  journal={IEEE J. Sel. Areas Commun.}, 
  title={Learning Task-Oriented Communication for Edge Inference: An Information Bottleneck Approach}, 
  year={2022},
  volume={40},
  number={1},
  pages={197--211},
  doi={10.1109/JSAC.2021.3126087}
}

@misc{isaac_sim, 
 title={{What is Isaac Sim?}}, 
 howpublished={\url{https://docs.omniverse.nvidia.com/isaac-sim/latest/index.html}}, 
 journal={Omniverse Isaac Sim Latest Documentation},
 note ={(accessed Feb. 2024)},
}

@misc{jackal, 
 title={{Jackal: Unmanned Ground Vehicle}}, 
 howpublished={\url{https://clearpathrobotics.com/jackal-small-unmanned-ground-vehicle/}}, 
 journal={Clearpath Jackal UGV Documentation},
 note ={(accessed Jul. 2024)},
}

@inproceedings{NIPS2017_d8bf84be,
 author = {Qi, Charles Ruizhongtai and Yi, Li and Su, Hao and Guibas, Leonidas J},
 booktitle = {Adv. Neural Inf. Process. Syst.},
 editor = {I. Guyon and U. Von Luxburg and S. Bengio and H. Wallach and R. Fergus and S. Vishwanathan and R. Garnett},
 pages = {},
 publisher = {Curran Associates, Inc.},
 title = {PointNet++: Deep Hierarchical Feature Learning on Point Sets in a Metric Space},
 url = {https://proceedings.neurips.cc/paper_files/paper/2017/file/d8bf84be3800d12f74d8b05e9b89836f-Paper.pdf},
 volume = {30},
 year = {2017}
}

@ARTICLE{9830752,
  author={Xie, Huiqiang and Qin, Zhijin and Tao, Xiaoming and Letaief, Khaled B.},
  journal={IEEE J. Sel. Areas Commun.}, 
  title={Task-Oriented Multi-User Semantic Communications}, 
  year={2022},
  volume={40},
  number={9},
  pages={2584--2597},
  doi={10.1109/JSAC.2022.3191326}
}

@ARTICLE{lsathomas,
author = {Thomas K Landauer, Peter W. Foltz and Darrell Laham},
title = {An Introduction to Latent Semantic Analysis},
journal = {Discourse Processes},
volume = {25},
number = {2--3},
pages = {259--284},
year = {1998},
publisher = {Routledge},
doi = {10.1080/01638539809545028},
}

@article{shannon1948mathematical,
  title={A mathematical theory of communication},
  author={Shannon, Claude Elwood},
  journal={Bell Syst. Tech. J},
  volume={27},
  number={3},
  pages={379--423},
  year={1948},
  publisher={Nokia Bell Labs}
}

\end{document}